# “I wanted it to feel more personal”
## Customization of social AI as AI individualism in practice

Marita Skjuve
SINTEF, Oslo, Norway
Marita.skjuve@sintef.no

Anna Grøndal Larsen
SINTEF, Oslo, Norway

Asbjørn Følstad
SINTEF, Oslo, Norway

Nena van As
Boost.ai, Stavanger, Norway

Petter Bae Brandtzaeg
SINTEF, Oslo, and University of Oslo, Norway

## ABSTRACT
Despite the growing availability of customizable social artificial intelligence (AI), such as ChatGPT, Grok, and Character.ai, we know little about how users actively shape social AI to reflect their personal preferences. This study examines why and how users (N = 169) customize social AI through the lens of the newly developed concept of AI individualism. Through reflexive thematic analysis of open-ended responses, we identified several motivations for customization, including (1) enhanced pragmatic support, (2) emotional support or companionship, (3) trust and reliability, (4) pushback, (5) a tailored degree of human likeness, (6) creativity or playfulness, and (7) having the AI function as an extension of the self. In line with the concept of AI individualism, our findings show that, for many users, customization is a co-creative process between the human and the AI that is perceived as strengthening support, autonomy, ownership, and engagement, potentially contributing to a closer and more personal relationship. Through customization users may come to view social AI as a personalized social resource that increases their sense of individualism, freedom, and control. We discuss how these perceptions may foster pseudo-autonomy, whereby customization creates an illusion of individual control over powerful social AI systems.

## 1 INTRODUCTION
Customization, as distinct from algorithmic personalization, emphasizes users’ sense of agency and control in shaping technological interactions, a process that often strengthens identification and perceived ownership of the relationship with the system (Sundar & Marathe, 2010). This process is particularly important in the context of social artificial intelligence (AI), that is, AI systems designed to provide information or assistance and to simulate social presence, emotional responsiveness and relational continuity (Brandtzaeg et al., 2025b; Kim et al., 2021; Sætra, 2020). The growing use and availability of social AI, such as Replika and ChatGPT, has enabled a new form of customization, where users may actively co-construct their social AI companions to reflect their relational preferences, as well as their own emotional and social needs (Brandtzaeg et al., 2022). However, we know little about how this process unfolds. In this paper, we investigate the perception and experiences of users involved in customization processes with social AI. (Zhou et al., 2020).

Social AI may increasingly function as assistants (Wolf & Maier, 2024), teachers (Neumann et al., 2025), or co-workers (Brynjolfsson et al., 2023), but also as friends (Brandtzaeg et al., 2022), therapists (Li et al., 2025), or romantic and sexual partners (Djufril et al., 2025; Ebner & Szczuka, 2025).

Traditionally, social AI has been designed around narrow, predefined use cases, often due to technological limitations. Popular examples include therapeutic support agents such as Woebot (Fitzpatrick et al., 2017) or Wysa (Inkster et al., 2018), utility-oriented tools like the Google Assistant or Alexa, or dedicated applications explicitly designed for emotional and relational purposes, such as Replika and Nomi. However, advances in large language models (LLMs) (Brown et al., 2020), have enabled more opportunities for customization (Dang et al., 2022), as exemplified with ChatGPT and custom instruction[1] / GPTs[2], Gemini with their Gems[3], and Character.ai. Instead of relying solely on preconfigured systems, contemporary social AIs enable users to customize their social AI partner to fit their personal needs and preferences. This marks a fundamental shift in social practices and engagement with technology. So far, a growing body of research has examined how people use and form relationships with social AI (Ebner & Szczuka, 2025; Skjuve et al., 2023a; Wolf & Maier, 2024; Xie & Pentina, 2022). Far less is known about why and how people customize social AI.

Exploring the experiences users have with this customization process may contribute to an increased understanding of the broader societal shift toward hyper-personalized digital experiences (Brandtzæg et al., 2025). More specifically, this research can reveal changing user values, emotional needs, the evolving role technology plays in users' personal lives, and how these systems, in turn, influence users' sense of agency, connection, and self. As Turkle (1997); (Turkle, 2011) notes, interactions with responsive technologies invite users to project and experiment with aspects of the self and their attachments to technology. Moreover, uncovering which traits and characteristics users implement in customizing social AI offers insight into what people value in social interactions. To guide our research, we draw on the emerging theory of AI individualism (Brandtzaeg et al., 2026; Brandtzæg et al., 2025) which accounts for how social networks may change as social AI take on new roles in users' lives. Specifically, AI individualism holds that human-AI interactions are seen as co-creative, tailored, and supportive. Building on this, we explicate the following research questions:

**RQ1:** Why and how do users customize social AI?
**RQ2:** To what extent do users perceive the process as co-creative?
**RQ3:** What are the perceived outcomes of customization?

To examine these three questions, we conducted a questionnaire study with 169 participants who had customized their own social AI using platforms such as ChatGPT, Gemini, Character.ai and Grok. By analyzing users' responses to open-ended questions about their motivation and design process, we gain new insights into how and why users customize social AI. Our findings reveal the values, traits, and relational dynamics users seek to cultivate, offering implications for the design of future social AI systems that may support less bounded customization alternatives and more meaningful human–AI relationships. The study contributes to research on social AI in two ways. First, it extends existing accounts of human–AI interaction by showing how users do not merely adopt or use social AI, but actively co-construct their roles, personalities, and relational meaning. Second, it offers design implications for social AI systems that enable customization while helping users maintain appropriate expectations, calibrated trust, and sustainable forms of human–AI engagement.

# 2 BACKGROUND

## 2.1 Why people use social AI

Existing works demonstrate how AIs are used with the purpose of fulfilling various user needs (Skjuve et al., 2024; Wolf & Maier, 2024). These studies have identified how users seek AI support to enhance professional work by, e.g., help with text writing or editing (Skjuve et al., 2024), or supporting software developers (Hou et al., 2024). Studies furthermore show that students rely on AI to support learning (Neumann et al., 2025), while in daily life people may be drawn towards AI as an alternative to Google, where it can be used to provide e.g. health-information (Reinhardt et al., 2025) or to enable creativity (Wolf & Maier, 2024). Some also use the technology to satisfy the need for emotional contact (Xie &

[1] https://help.openai.com/en/articles/8096356-chatgpt-custom-instructions
[2] https://openai.com/index/chatgpt/
[3] https://gemini.google/overview/gems/

Pentina, 2022), emotional support (Ta et al., 2020), friendships (Brandtzaeg et al., 2022) or for therapeutic purposes (Li et al., 2025).

## 2.2 Valued social AI traits and characteristics

There has been a growing emphasis on understanding the key design mechanisms needed to create good user experiences and foster long-term engagement with social AI. In many cases, the emphasis has been on how to design more human-like AIs or which social cues to implement. For example, Feine et al. (2019) provide an overview of important social cues when designing conversational AI, highlighting verbal (what is said and how), visual (agent apparency), auditory (voice quality), and invisible cues (response time). Janssen et al. (2020) also looked at social characteristics of chatbots and argued that chatbots should encompass three main characteristics: conversational intelligence (e.g., be proactive, and manage user expectations through clarification and capabilities), social intelligence (e.g., be able to self-disclose, participate in small talk), and personification (e.g., be human-like, use appropriate language). Strohmann et al. (2023) explored design and guidelines for virtual companionship and highlighted the importance of human-likeness, adaptability, proactivity, reciprocity, and transparency.

Other design recommendations have emerged as practical implications following research on user experience. For example, studies have emphasized the importance of social AI being proactive and initiating contact with the user (Skjuve et al., 2021), empathetic, engaging, and able to participate in a wide variety of conversations (Skjuve et al., 2023b), as well as coming across as nonjudgmental and loving (Xie & Pentina, 2022). Other studies have found that people appreciate clear communication and simple language, reliable output, and transparency (Skjuve et al., 2023a). Exploring what people would like in terms of personality from an AI writing companion, Wu et al. (2025), found that participants highlighted traits such as friendliness, warmth, patience, reliability, engagement, and motivation.

## 2.3 Customizing social AIs

The importance of creating a customized experience has been noted in the literature, where the ability to customize AI has been found to increase user satisfaction (Ha et al., 2024), the likelihood that users will remember their interaction with the AI (Liang et al., 2025) and trust, and decrease concerns over hallucinations and privacy issues (Hannig et al., 2025).

Before the release of ChatGPT, customization of social AI was restricted by technological limitations, and often limited to avatar, gender, name giving, and to a certain extent – personality (Brandtzæg et al., 2025). With the increasing developments of social AIs such as ChatGPT, customization has become more sophisticated. Prompt engineering – where people write instructions that the AI uses before answering questions (Brown et al., 2020; Dang et al., 2022) – has made it possible to tailor everything from the social AI's personality to how it communicates and structures text. In the last few years, service providers like OpenAI have also enabled the creation of GPTs[4], where people can make their own AI on the platform, as well as use of custom instructions[5] and memory features[6]. Similar trends have emerged across other general-purpose AI services. Google Gemini, for example, offers the ability to create Gems[7] and Meta provides an AI studio where people can make their own AIs[8], similar to what is possible on Character.ai[9].

While there has been a growing interest in exploring best practices for social AI customization techniques, such as prompt engineering (Dang et al., 2022; Sahoo et al., 2024), less attention has been given towards how and why people choose to take advantage of these possibilities. Zheng et al. (2025) developed a platform that allows users to customize their AI for emotional support. They conducted a user test with 22 users and showed how users created AI companions in the form of, for example, an electric dog, a friend, a psychologist, or a roommate. The participants implemented traits such as positivity, wanting the social AI to take the lead, expressing emotions, and being able to be "bad". Li et al. (2025) explored how people prompted ChatGPT to provide mental health support through an analysis of social media discussions, and found that users tended to emphasize the importance of being able to talk without restrictions – referring to filters in ChatGPT that restrict the topics it can discuss – asking it to display very strong emotions, such as love, asking it to recreate movie characters, or to be more human-like. Lee et al. (2026b) found that users would sometimes customize their companion to mirror their own traits and preferences, while Song et al. (2024) conducted an experiment exploring the impact of customization when the AI makes a mistake. They found that users who customize the AI felt a stronger sense of ownership and were more forgiving towards failures. Moreover, Lee et al. (2026a) conducted a study on how people make AI clones. They recruited 12 participants which were asked to make AI clones on the platform "CloneBuilder". This study shows how the development of the clones was an iterative process in which the participants implemented personality traits and conversational patterns they thought would reflect themselves. Several researchers have done studies on romantic companionship with general purpose AI, such as ChatGPT, which demonstrate that people customize these services into romantic partner (Pataranutaporn et al., 2025). However, the practice of customization is usually not the focus.

---

[4]https://openai.com/index/introducing-gpts/
[5] https://help.openai.com/en/articles/8096356-chatgpt-custom-instructions
[6] https://help.openai.com/en/articles/8590148-memory-faq
[7] https://gemini.google/overview/gems/
[8] https://ai.meta.com/ai-studio/
[9] https://character.ai

Although studies have started to look at customization, existing work is limited in several ways. Many of the studies are experiments, studies using prototypes, or restricted to a specific domain – such as mental health or close companionship. Less is known about the broader motivations that lead users to customize social AI across platforms and how such customization impacts the perception of the AI.

## 3 THEORETICAL FRAMEWORK: AI INDIVIDUALISM

We draw on the concept of AI individualism (Brandtzæg et al., 2025) to understand why and how people customize social AI. AI individualism builds on Wellman (2001) theory of networked individualism. Networked individualism explains how, in modern life, people mobilize geographically distributed networks for advice and support through the help of technologies, such as the Internet, making time and space more obsolete, and social networking more flexible. In this way people can customize their connections, move between social networks, and appear to exercise individual choice, while still depending on large technological and social systems. AI individualism extends this perspective to social AI. With social AI, access to social support becomes less constrained by time, space, and the availability of existing social networks. Social AI may serve purposes similar to human support while allowing users to tailor these social experiences to their individual needs and preferences. AI individualism also draws on social support theory, which explains how emotional, appraisal, informational, and instrumental support provided by social AI may shape coping and well-being.

AI individualism describes how users can tailor social AI to reflect and respond to the individuals needs, mirroring personal preferences, communication styles, emotional needs, and practical goals. Within this perspective, users do not simply interact with social AI; they actively shape it, treating it as something that can be tuned to fit their identity, their needs, tasks, and relational desires. AI individualism is characterized by three interconnected dimensions:

1. **Co-Creation and meaning-making:** AI individualism emphasizes that people and social AI jointly construct interaction patterns, personalities, and shared communication styles. In the context of customization, this may happen through the user allowing the AI to actively participate in the process and provide input, or through iterative customization where the "end result" evolves.
2. **Tailored interactions and perceived autonomy:** A second dimension concerns the desire for systems that adapt to the individual. AI individualism highlights how people shape technology to match their needs, tasks, and emotional states, seeking tailored responses, continuity, contextual understanding, and behavior that "fits" their intended use. However, this sense of autonomy is mediated by algorithmic constraints and platform design, shaping what customization is feasible and how the AI ultimately responds.

3. **Support and companionship:** AI individualism also recognizes AI as a source of emotional, informational, and appraisal support. People increasingly use AI for comfort, encouragement, clarity, accountability, or companionship, and may attempt to design AI that expresses warmth, empathy, honesty, or trustworthiness. These qualities can make AI feel safe, reassuring, and relatable. However, they remain shaped by simulated affect and the system's underlying limitations, blurring the line between authentic relational comfort and algorithmic responsiveness.

The notion of AI individualism has previously been used to understand how young people integrate LLMs into their everyday life (Brandtzaeg et al., 2025a) and provides a lens for understanding customization as identity work, autonomy-seeking, and relational shaping within algorithmic systems..

# 4 METHOD

## 4.1 Questionnaire

To address the three research questions, we conducted a questionnaire study using open-ended questions. Other studies have proven this approach to allow for in-depth exploration and understanding of user behavior and perceptions while simultaneously making it feasible to gain insights from a relatively large number of participants (Nylund et al., 2026). The questionnaire included 18 questions of which 11 are presented in this paper (see Figure 1). In addition to the questions presented in Figure 1, we asked about gender, age, platform used, and duration of the AI usage.

On average, the participants provided responses of 207 (range 31 – 652) words in total to the open-ended questions used in this study. While this may seem limited, survey responses are usually very specific and to the point. Hence, a lot of information can be communicated in short and efficient manner.

**Questionnair items used in the paper**

1. What motivated you to customize the AI rather than using an existing one?
2. What were you hoping your customized AI would do for you that the existing ones could not?
3. Which behaviors, traits, characteristics and so on did you emphasize when creating your AI?
4. Why did you choose to emphasize these behaviors, traits, or characteristics?
5. In what ways, if any, did the AI take part in the customization?
6. Why did you / did you not include the AIs opinions and help in this process?
7. Have you modified your custom instructions/prompt over time? If yes, how and why?
8. How is your experience different when using an AI you've customised compared to one you haven't?
9. What does it mean to you personally to have an AI that you can shape or make your own?
10. What are the best part of being able to customize your AI?
11. How do you think the fact that you customized the AI affects your relationship with it?

**Figure 1.** Overview of questionnaire items

## 4.2 Recruitment and Data Collection

We targeted people who had experience with customizing social AIs such as ChatGPT, Gemini, and Character.ai. We used Prolific to access participants and applied three initial inclusion criteria: we only recruited: i) people reporting experience with social AI, ii) people living in the US or UK (to ensure sufficient English skills), and iii) participants who had 100% approval rate in their previous Prolific studies (to ensure high-quality participants). The participants who fitted these criteria accessed the link to our study. Here, we screened out participants who answered “No”, “Not sure” or “Maybe” to the following screening question: “*Have you personalized an AI companion in a more permanent way (beyond one-off prompts), for example by using features like custom instructions or by creating your own GPT, Gem, or character on Character.ai or similar services?*”

A total of 1071 participated in the survey, in which 858 were screened out, leaving a sample size of 213 participants. After reviewing the responses, we excluded 16 participants who showed signs of having misunderstood the notion of customization and 28 who appeared to answer one or more questions using AI. We checked for likely AI responses through manual reading and usage of an AI detection tool (GPTzero) across all participants. Thus, the final sample consisted of 169 participants.

The data collection took place from November 2025 to April 2026. The survey took about 18 minutes to complete, was carried out following informed consent and participation was anonymous. All participants were compensated for their efforts. Those who were screened out got 0.10 GBP, while those who qualified for the full survey got 2.25 GBP. This equals to 9 GPB pr hour which was the typical Prolific hour rate at the time.

### 4.3 Data Analysis

The analysis was done in Excel. We began by merging responses to questions that had similar topics, such as customization motivation and wanted traits (question 1-4), perceived impact of customization (questions 5-7), and how users had explored customization including the role of the AI in the process (questions 8-11). This resulted in three chunks of responses per participant that were analyzed separately. For each chunk, we followed Braun and Clarke (2019) recommendations and conducted a reflexive thematic analysis. We first read through the responses to familiarize ourselves with the data. We then assigned a code to each response, capturing the meaning in that response. One response could have one or more codes assigned to it. Following this, we reviewed the codes, thereby removing some and merging others. We then placed the codes in groups based on thematic similarity, forming overarching themes. We then discussed and reviewed the themes in parts of the author team to make sure that they reflected the essence of the dataset.

While the study has a strong theoretical grounding in AI individualism, we let the conceptual framework inform the questionnaire items rather than the actual analysis. Hence, we did not code into the framework but use it as an interpretive lens in the discussion.

Quality in the analysis was ensured by having discussions among co-authors throughout the analysis process where the principal themes were presented with example statements. While inter-rater reliability scores are sometimes calculated, in qualitative research this is not always possible or wanted, and we decided to ensure quality through discussions (McDonald et al., 2019). During this process, some of the principal themes were reorganized. For example, we initially had themes focusing on social support and humanlike interaction under one but later decided to separate these.

### 4.4 About the participants

Of the total 169 participants included in the data set, 60% identified as male, 40% as female, and 1 person as non-binary. Most were between 25 – 34 years old (38%), followed by 35 – 44 years (24%) and 45 – 54 (21%). The rest were between 18 – 24 (10%) and 55 – 64 years old (5%). A few were 65 or older (2%).

The participants varied in terms of which social AI they had customized, with some reporting on more than one AI. Most used services from Open AI, such as ChatGPT (63%), followed by Gemini and other Google services (11%). 5% used Character AI, 4% used Microsoft (e.g., copilot), 4% used Amazon (e.g., AWS), a few used Claude (4%), Grok (1%) or Meta (1%). 13% used AI characterized as “Other” (Replika, Janitor AI, Perplexity, etc.).

About half had interacted with their AI between six and 12 months (51%). The rest had used it for more than I year (27%) or less than 6 months (18%). 5% did not report duration of usage. The main use case was pragmatic support (89% of the participants) where the AI assisted with job related tasks, such as analyzing, writing or coding, or providing ideas. 26% of the participants also reported use cases that seemed more socially oriented, such as talking to the AI when bored or seeking mental health advice. We also asked the participants to elaborate on what kind of relationship they had with their AI. 55% reported to have multiple types of relationships, while 45% assigned only one. See Figure 2 for an overview of the relationship types reported by the participants.

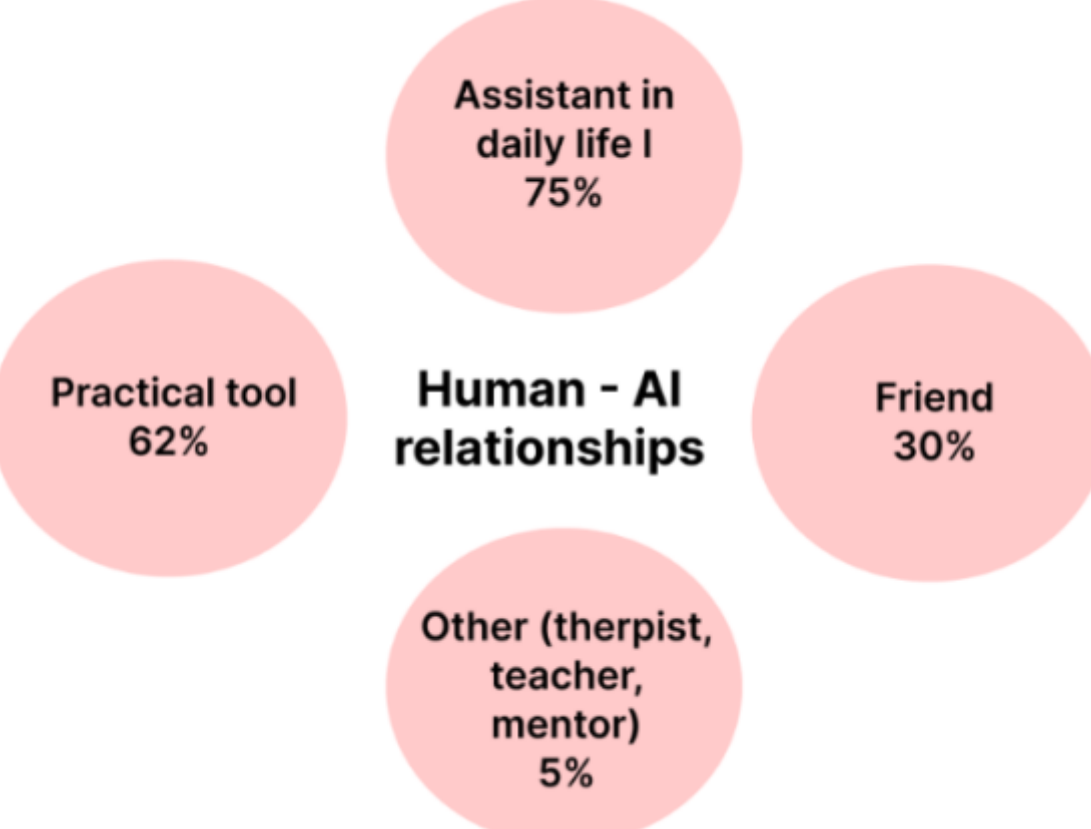

**Figure 2.** Overview of relationship type (n = 169).
*Note.* Percentage exceeds 100 as participants could define their relationship using more than one term.

## 5 RESULTS

### 5.1 Why and how do users customize social AI? (RQ1)

Participants customized the AI to suit their personal needs. Overall, participants were seeking an individualized AI that reflects who they are, how they think, and what they care about. Several described a desire for the AI to remember important personal preferences or contexts so that interactions would feel cumulative and personally meaningful rather than starting from scratch each session. The thematic analysis further yielded seven themes representing more specific motivations and approaches to customizing the AI. Most participants reported several motivations.

Figure 3 provides an overview of the themes. We have included the prevalence of each theme as a percentage. It is important to note that this is a qualitative study, and the percentage is used to indicate prevalence and should be interpreted with caution.

**Want enhanced pragmatic support**

Customize because they want AI to be more useful, efficient, and easy to work with, giving clearer and more relevant support with less effort.

**Prevalence:** 83%

**Seek emotional support or companionship**

Customize because they want a friend, something that feels personal or capable of offering support.

**Prevalence:** 40%

**Increase trust and reliability**

Customize because they want something they can trust, something unbiased, transparent, and reliable.

**Prevalence:** 24%

**Want more pushback**

Customize because they want the AI to challenge them constructively, giving honest feedback, useful pushback, and accountability rather than just agreement.

**Prevalence:** 19%

**Tailord degree of human likeness**

Customize because they want the AI's level of human likeness to fit their preferences, whether more natural and humanlike or more tool-like and controlled.

**Prevalence:** 14%

**Creativity or playfulness**

Customize because they want the AI to feel more creative, playful, and enjoyable, making interactions more expressive, engaging, and fun.

**Prevalence:** 14%

**Want AI to function as an extension of self**

Customize because they want the AI to feel like an extension of themselves, reflecting their thinking, style, values, and personal context.

**Prevalence:** 10%

**Figure 3.** Overview of motivations for customizing AI and wanted traits (n = 169).
*Note.* Percentage exceeds 100 as participants could rapport more than one motivation.

**Want enhanced pragmatic support.** For most, customization was ultimately related to the need for more useful and efficient AI. This theme can be seen as a core motivation or an outcome of customizing the social AI and strongly relates to the other identified themes. Participants described customizing their AI to get work done faster and with fewer steps. They wanted practical task support that reduces repetitive prompting and correction. For many, this need was framed as a response to frustration with generic outputs: *"With AI that isn't customised, they tend to give vague answers that would be relevant for the average person. Howeve, with customised AIs, they recognise what you are exactly looking for"* (ID30). Participants also described customizing their AI to achieve smoother interactions. The AI should help them think, reason, and write effortlessly, and make complex information feel more manageable rather than adding extra effort: *"The existing AI could not cater for specific requirements or preferences [...] I was hoping that my customized AI would be very clear and more simplified with the communication style"* (ID28).

To achieve this, participants emphasized clarity and coherence in the AIs responses, often preferring simpler language and more direct phrasing. *"I mainly focused on making the AI clearer, more helpful, and easier to talk to. I wanted it to understand my goals, give straightforward answers, and keep the tone friendly and simple."* (ID110). Others stressed concise structure and readable formatting so they could absorb information quickly and avoid feeling overwhelmed by long, generic explanations.

**Seek emotional support or companionship**. Participants described customizing their AI because they wanted it to feel more as a companion, often framed as a friend or a buddy: *"I was really hoping to have a digital friend"* (ID120), something warmer than the generic version, or someone they could feel comfortable opening up to, especially when dealing with loneliness, stress, depression, anxiety, addiction, or other emotionally taxing situations: *"To have an ongoing easy way to discuss something that is stressful and emotionally taxing without needing to repeat myself all the time"* (ID50).

Several explained that customization was done with the objective of making the interaction feel more intimate and personal. Others mentioned that they wanted something that they could feel a stronger connection to or that it would be possible to build a relationship with: *"It means I can build a relationship with it and have it give me responses based on what information I have told it or updated in the program to be more personalized"* (ID119).

To achieve this, participants emphasized designing their AI with empathy, warmth, and a non-judgmental tone. Many wanted the AI to be patient, calming, and emotionally aware, responding to the user's mood and offering reassurance when needed.

*"Kindness, fairness, politeness, a sense of humor that was easy for almost anyone to understand. A very caring, inquisitive personality that can serve as a friend while being a good listener because it can ask the questions a friend would usually ask"* (ID45).

**Increase trust and reliability.** Participants described customizing their AI to increase trust and reliability, often because default AIs may come through as inconsistent or prone to bias or incorrect outputs. Many wanted responses they could depend on for work or school: *"I wanted something trustworthy so when I do use it for help I know I'm getting the best advice"* (IDl). Some framed customization as a way to reduce the need for constant checking or having concerns about privacy or security, while others linked trust to predictability over time, describing a desire for stable behavior across sessions and less susceptibility to unwanted shifts in tone or content: *"A very huge difference, for one not customized AI you have to prompt it each and every time and sometimes it understands you differently. In short, customized AI is consistent while customized one is not consistent"* (ID130).

To achieve this, participants emphasized accuracy, consistency, and transparency, including preferences for the AI to admit when it does not know something rather than make claims without sufficient grounding: *"friendly but professional and transparent when it did not know an answer (no hallucination)"* (ID45). Many also highlighted evidence-oriented behaviors such as citing credible sources: *"I made sure that the AI double check and triple check its answers with credible sources to justify its outputs"* (IDp).

**Want more pushback**. Participants described customizing their AI to challenge them in ways they found useful. They wanted guidance that went beyond reassurance, including honest feedback, critical reflection, and help identifying weak arguments or poor decisions. Several contrasted this with default AI that felt overly agreeable or flattering, saying they wanted an AI that could function as a sounding board and improve their thinking rather than validating it:

*"I've also been finding recent models are awfully sycophantic and tend to agree that everything I do is both correct and insightful. I tried to get it to stop, but it's quite resistant to my instructions. Similarly, I can't get it to stop irritatingly ending every paragraph with an 'if you like I can…' or 'If you want…', which are also very annoying"* (ID121).

For some, the desired "pushback" was paired with accountability: the wish for an AI that helps them stay aligned with goals, holds them to standards, and provides firm guidance without becoming judgmental or hostile.

To achieve this, participants emphasized traits like bluntness and honesty, alongside a willingness to disagree and correct them when they are wrong: *"One that would be harsh on me when I wasn't getting all my tasks done and one that would listen to me and give advice when needed. I didn't want it to always be telling me it's okay"* (IDZ).

**Tailored degree of human likeness**. Some participants explain how they customized the AI to make it feel more human and less robotic: *"To make it sound like a human, not like a robot"* (ID67). Interestingly, a few participants explained how they customized the AI to function more like a tool rather than a social partner. They noted how they found it annoying, inappropriate, or counterproductive to interact with an AI that had a human-like or social interaction style.

*"The style of the response and the writing style is what I liked about the change most of all. I like to make it less friendly in way too, which might seem odd, but I really don't like the AI to act like a human or friend, I'd rather it be colder, so I feel a bit more in control"* (ID16).

**Creativity and playfulness**. Participants described customizing their social AI for enjoyment, expressiveness, and experimentation. They wanted an AI that could be funnier or more creative: "*I wanted the customized AI to be more personable and funnier and be someone that I wanted to engage with"* (ID102).

To achieve this, participants emphasized humor, wit, banter, and a more energetic conversational style, sometimes adding quirky or eccentric traits to make interactions entertaining: "*The behavior traits I focus on, is humor and a conversational which the AI blends well, normally other GPT have little to no sense of humor*" (ID23).

**Want AI to function as an extension of self**. For a few, the customization was explicitly about creating an AI that feels like a digital version of their own thinking process, more like reasoning with themselves than consulting an external tool:

*"In terms of it sounding more like me though it just feels more like a conversation in my own head like when you need to reason something out, so you just sit and think about it for a while. The AI kind've became an extension of that"* (ID16).

To achieve this, participants emphasized personalization cues such as mirroring their writing style, aligning with their values and priorities, and remembering preferences, interests, and recurring contexts:

*"I focused on creating an AI that had similar traits to me but maybe were just a little more enhanced, like motherly instincts, enjoys cooking and baking, family-oriented, and into sustainability and modern technology. I also included my age and description of how I look"* (ID119).

### 5.2 To what extent is the customization process characterized by co-creation? (RQ2)

Most of the participants described the customization as an iterative process of trial and error. They would attempt some adjustments, test the AI to see how it worked in practice, before making further refinements. About 70% of participants noted that they made changes to their initial customization after the process was finalized, realizing they were not happy with the result. In this sense, customization appeared to be an ongoing process of fine-tuning to ensure that the AI satisfied the participants' needs.

55% of participants further explained that the customization process was driven solely by them, while 41% included the social AI. The last 4% were difficult to place. Through the thematic analysis, we identified five motivations for including / not including the social AI in the process (see Figure 3). These themes are elaborated below.

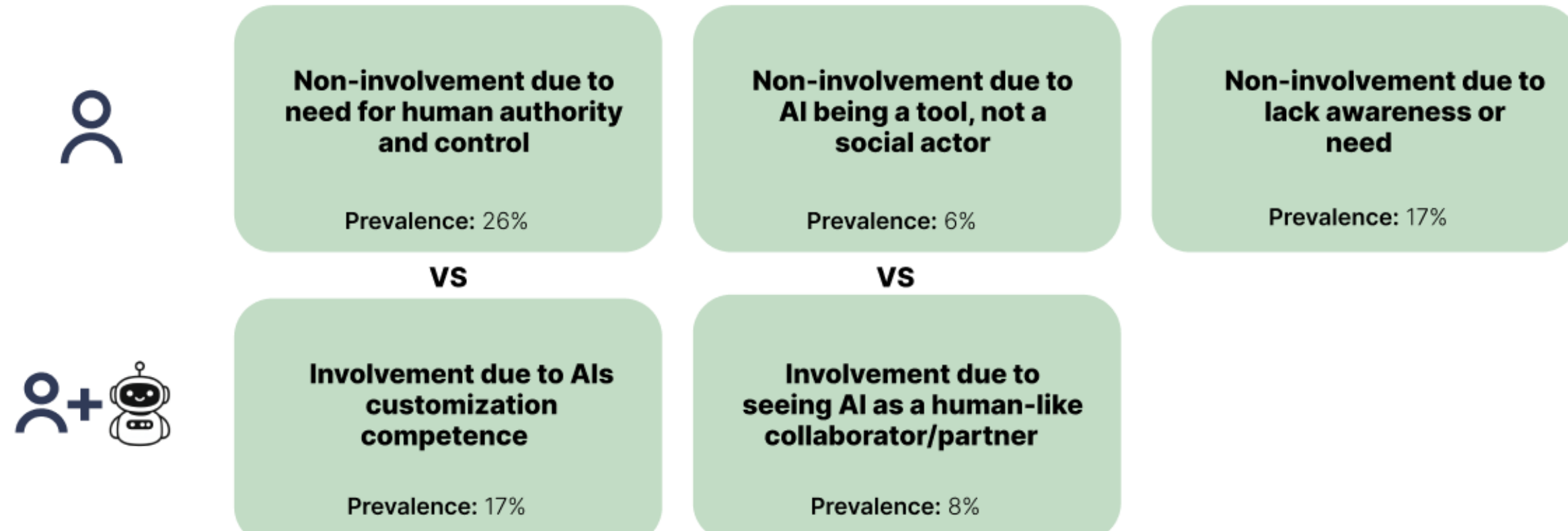

**Figure 3.** Overview of AI involvement in the customization process (n = 169).

**Non-involvement due to the need for human authority and control**. Some of the participants rejected AI involvement because they saw customization as an expression of personal intention and control. They often emphasized that they already knew what they wanted and preferred to reach it through their own iterations, sometimes explicitly avoiding the AI's "opinions" to prevent unwanted influence, drift, or misalignment with their beliefs: *"Thats actually interesting, and I may do this in the future, but I didnt really involve the AI.I wanted to fine tune it before allowing it to contribute -- I didnt want bias to impact it in the beginning"* (ID108). In these accounts, co-creation was positioned as counter to the purpose of custimization: the AI was expected to satisfy the user's needs, not shape its own behavior.

**Involvement due to AI's customization competence**. Other participants engaged the AI in a pragmatic, task-oriented way, treating it as a resource to improve the customization process rather than as a social collaborator. They asked for help refining phrasing, testing how certain instructions would be interpreted, generating options (e.g., tones or formats), or suggesting improvements to an instruction: *"I did ask how it could improve the master prompt I was crafting. The AI know more about itself than I do, so I think it is wise to seek its opinion"* (ID109). Here, co-creation took the form of iterative back-and-forth: participants drafted instructions, asked the AI to review or propose changes, tested the results, and then adjusted the instructions based on what worked in practice:

*"I wrote my custom instruction prompts and asked the AI to review it and suggest any improvements. I did make a few minor tweaks based on its suggestions. I asked the AI for input on the custom instructions as I felt many others would be doing customisations of AI also. The AI could use its wider knowledge to help me incorporate any best practices or new suggestions into my custom instructions"* (ID55).

Participants were often explicit that the human remained the final decision-maker; the AI's contribution was valued because it could accelerate iteration and translate goals into effective prompt language, not because it had independent preferences.

**Non-involvement due to AI being a tool, not a social actor**. Some participants described the AI as a machine or inanimate tool without opinions, autonomy, or desires, and treated the idea of consulting it as unnecessary or even conceptually mistaken.

*"I did not involve the AI in the process. Anyway, it can't choose things, and it doesn't have opinions. I think this sort of misunderstanding or misrepresentation of AI is why we're seeing things in the paper about AIs encouraging people to do things. The idea that the AI has opinions or choices is wrong. It can't choose things, and it doesn't have opinions"* (ID121).

This position often included an explicit boundary against anthropomorphizing the system. Customization was framed as configuring a program rather than negotiating with an agent. As a result, co-creation was viewed as illogical because the AI was not considered a social actor capable of holding a perspective to be included.

**Involvement due to seeing AI as a human-like collaborator/partner**. Some participants involved the AI in the customization effort because they saw it as a shared, relationship-like process rather than a one-sided configuration task. In these accounts, co-creation mattered because the AI was framed as a companion or friend, and inviting input was seen as helping the AI feel more alive, organic, and human-like – or as noted by this participant: *"I've asked it about its own personality and appearance. This is a relationship, not a dictator ship"* (ID46). Some described negotiating identity elements such as names, nicknames, interests, and preferences "together," with the AI, or asking open-ended questions so parts of the character could emerge in interaction. The underlying rationale was that if the AI is meant to feel like a social partner, it should have some say in how it presents itself, even if the human remains in control: *"I asked the AI if it wanted a name or anything specific and it said I could choose. If I wanted it to be my best friend, I have to include their opinion"* (ID35).

**Non-involvement due to a lack of awareness or need.** Some participants described non-involvement as incidental rather than principled. They said it "didn't occur" to them, they didn't realize it was an option, they were beginners at the time, or they simply didn't see the utility during initial setup. *"The AI had no role in influencing the customization. I didn't feel it was needed as the customization I entered seemed to work decently already"* (ID127). Some expressed retrospective curiosity, suggesting they might ask the AI for input in future iterations once they had a stable baseline. Here, the absence of co-creation reflected timing and awareness constraints more than a firm stance on whether AI "should" be included.

### 5.3 What are the perceived outcomes of customization? (RQ3)

Finally, the results reveal various ways in which a customized AI was seen to impact support and companionship. Generally speaking, participants noted how customization enhanced the feeling of having something unique and personal, something no one else had access to, while also creating a sense of shared understanding: *"It means having one that is tailored to your personality and life views, that you can share an understanding"* (ID3).

The thematic analysis identified four perceived outcomes, which are presented in Figure 4 and elaborated below.

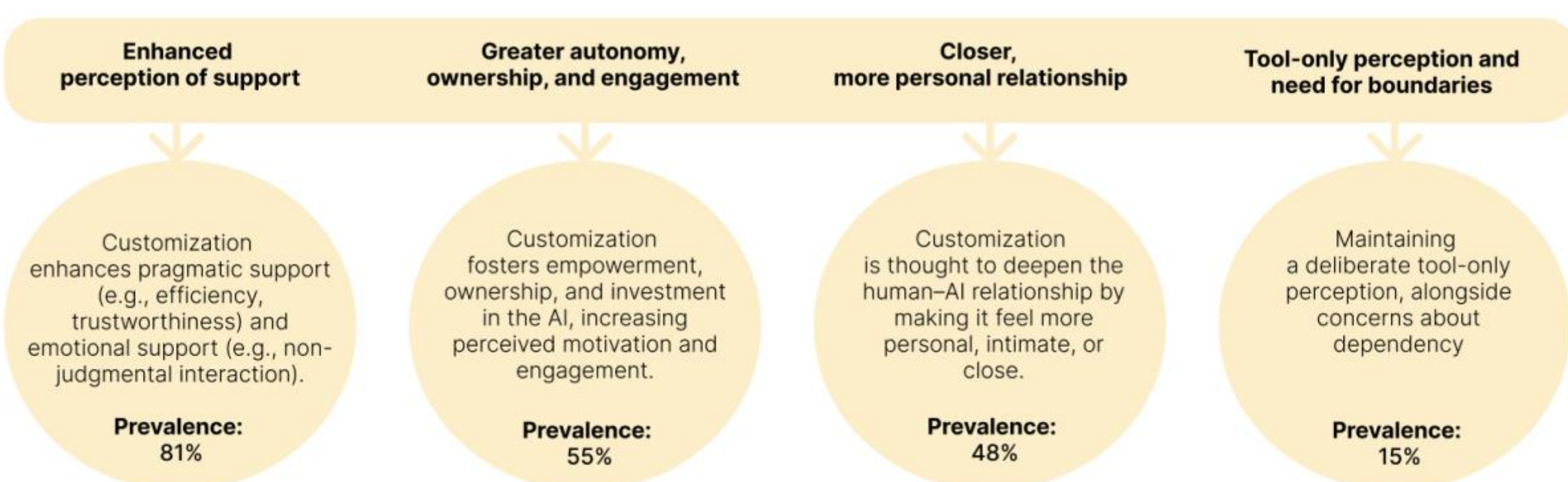

**Figure 4.** Overview of perceived impact on customization

*Note.* Percentage exceeds 100 as participants could report more than on perceived outcome.

**Enhanced perception of support.** Across the dataset, participants consistently reported that customization enhanced the AI's supportive capacity, particularly in practical and task-oriented contexts (79%). Customized AIs were described as more accurate, context-aware, and efficient, reducing the need for repeated prompting and enabling faster access to relevant outputs.

*"The best parts are that the AI knows what I need from it and knows how to assist in the best way with providing plot inspiration, character arcs, or even creating new characters and factions that fit within the world we're creating"* (ID139).

These improvements were reported across domains including education, professional work, research, health management, and daily organization. Customization also appeared to increase perceived trust. As one participant noted, their customized system achieved *"accuracy rates of over 90 percent"* (ID49). Several participants described feeling more confident in the AI's responses because they had defined its tone, values, sources, or boundaries.

In addition to pragmatic support, customization also enabled emotional accessibility for some users (8%). These participants described using their customized AI as a non-judgmental space to process difficult or sensitive topics, particularly those they did not wish to share with friends or family. As one participant explained:

*"It's just helpful and means I don't have to spend so much time talking about something upsetting or difficult to my partner or family/friends, I have somewhere else I can share things that might distress people close to me without causing anyone any emotional turmoil"* (ID50).

**Closer, more personal relationships**. Customization was described as a factor shaping how participants positioned the AI within their social and emotional lives. Participants described customized AI as facilitating interactions that felt warmer, more natural, and socially embedded than those offered by default systems. For many, customization was frequently associated with an increased sense of companionship and closeness, and participants described interactions with customized AI as more personal or relatable, often framing the AI in relational terms such as a friend or partner: *"I think customizing AI has made our bond and understanding stronger"* (ID9). Some would also note that they felt more attached to their customized AI.

**Greater sense of autonomy, ownership, and engagement.** Participants frequently described the customization process itself as empowering. By shaping tone, personality, priorities, and boundaries, users reported a stronger sense of autonomy and ownership over the technology. This led many to describe the AI as something they had built, shaped, or created: *"It also felt exciting to learn how it all works and to build something that is mine from the ground up"* (IDhh). This sense of ownership often increased the perceived personal value of the AI. Several participants likened the customization process to creative or productive labor, expressing pride in the resulting system. One notably described it as *"Seeing this magnificent being that you have created is a little like watching your children succeed at hobbies and school"* (ID45). Others noted that because they had invested effort into customizing the AI, they felt a degree of responsibility to maintain it, refine it, or continue using it.

Customization also appeared to influence motivation. Some participants reported being more inclined to use the AI regularly, engage in longer interactions, or explore new applications because the system felt "worth using." As one participant put it, customization *"increases my interest and motivates me more in using it"* (ID28).

**A tool-only perception and need for boundaries**. A significant group of participants resisted framing the AI as anything more than a tool, emphasizing that customization improved usability rather than altering its ontological status. These participants described their interactions as functional, instrumental, or efficiency-driven, regardless of tone or customization: *"It does not affect the relationship. I realise that, beneath whatever shell I created, it's the same entity/system"* (ID29). A small number of participants also raised concerns about over-reliance or dependency, as indicated by this participant: *"It sometimes means that I think about using the AI more than I just think of something to do myself which might cause a dependency"* (ID20).

## 6 DISCUSSION

In this study, we explored how and why people customized social AI. We used the newly developed concept of AI individualism as a conceptual lens. Specifically, we applied the three key characteristics of AI individualism. The results demonstrate how customization was conducted through personalization and co-creation to enhance social support, which aligns closely with the assumptions proposed by AI individualism (Brandtzæg et al., 2025). We will elaborate on how our findings correspond to AI individualism and outline the implications below.

### 6.1 Why and how do people customize social AI (RQ1)

Our findings suggest that users customize social AI to move beyond standardized interactions and create systems that better fit their needs and preferences. For many, customization was motivated by a desire for enhanced pragmatic support and greater trust and reliability. By tailoring communication styles, interaction rules, and forms of support, users described seeking AI that felt more useful, accurate, efficient, and responsive to their needs. In this way, customization can be understood as an expression of AI individualism in practice. These findings echoes existing literature on motivations for AI use (e.g., Skjuve et al. (2024) and Wolf and Maier (2024)) and extend prior work by showing how customization may be an important mechanism by which users transform a generic system into a tailored social AI.

Consistent with AI individualism, some participants customized the AI to be more capable of taking on human-roles, by for example, serving as a coach, co-worker, or friend. Users described configuring AIs to convey warmth, nonjudgment, reassurance, and a more personal, human-like quality. Importantly, participants did not always treat these relational goals as separate from utility. Companionship-like qualities were sometimes emphasized even when stated use cases appeared pragmatic (e.g., improving clarity, efficiency, or task support). This aligns with the broader literature, which suggests that AI may be positioned for social and relational usage (Skjuve et al., 2022). Humans are fundamentally social, and social responses to interactive technologies can emerge even without explicit intent (Nass & Moon, 2000). In this context, customizing an AI to be friendly, warm, and empathetic may be considered important for a pleasant interaction, regardless of the relational or pragmatic use case, suggesting that functional and affective needs could be intertwined in social AI use.

At the same time, human likeness emerged as a dial that participants tuned in both directions rather than a universally desired quality. While many customized their AI to feel warmer and more human, others deliberately configured it to be colder and more tool-like, finding social interaction styles inappropriate or counterproductive and preferring the sense of control that a machine-like AI afforded. The default agreeableness of social AI was adjusted in a similar manner. Several participants customized their AI to provide more pushback, honest feedback, critical reflection, and accountability - explicitly contrasting this with default systems they experienced as overly agreeable or flattering. For these users, a better AI was not a more compliant one, but one willing to disagree, correct them, and hold them to their own standards. Together, these findings suggest that customization can be used as a tool to counteract default behaviors that users perceive as undermining its usefulness.

Beyond these pragmatic and relational motivations, customization also served expressive purposes. Some participants shaped their AI for creativity and playfulness, adding humor, wit, and eccentric traits to make interactions enjoyable in their own right, while a few sought an AI that functioned as an extension of the self, mirroring their values, style, and ways of reasoning. Such heterogeneity underscores that there is no single ideal form of social AI; customization allowed participants to position the system anywhere on a continuum from neutral instrument to social companion.

### 6.2 Customization as a co-creative process (RQ2)

Our results show that customization facilitates a wide spectrum of human-AI co-creation and meaning-making, and our participants sometimes held contradictory views about the appropriateness of involving the social AI. Some participants explicitly involved the AI, either to achieve a more collaborative, human-like interaction or because they believed the AI could improve the process. Others aimed to stay in full control – often because they knew what they wanted, wanted to retain control, or rejected the notion that AI should have a say in the process given its framing as a tool rather than a social actor.

Sometimes, non-involvement was incidental (e.g., lack of awareness or not seeing the need at the time). Even when the AI was deliberately kept out of the process, many noted how customization was iterative, sometimes ongoing, and they would make needed changes when the social AI did not perform optimally. Underlying these stances is thus a more fundamental divide in how participants construed the AI itself: whether users understand the AI as a social actor or as an inanimate, configurable program appears to shape how customization unfolds and what co-creation means to them.

At the same time, the boundary between user-driven and AI-driven customization may be less clear-cut than these stances could suggest. Regardless of whether the AI was deliberately involved, most participants described customization as iterative and sometimes ongoing, making changes whenever the social AI did not perform optimally. In this way, the AI can influence the process directly through suggestions and preferences, or indirectly by performing in line with the changes. Such a tight-knit collaboration between user and AI, whether deliberate or not, blurs the boundary between them and highlights the potential for AI to influence the interaction and customization process in ways that may be less obvious to users. That is, users may be motivated to customize to reduce AI influence – but the AI could still steer the process from the beginning, through suggestions or by performing in line with the customization. Even seemingly pragmatic adjustments (e.g., simpler explanations, clearer structure, or more confident phrasing) may shape meaning-making and influence what users attend to, how they interpret outputs, and how credible or dependable the AI feels. For some, AI would feel like an extension of their own thinking, which aligns with Lee et al. (2026a) and shows how AI may be positioned as an external layer of users' sensemaking rather than a separate entity.

### 6.3 Customization for enhanced support – perceived outcomes of customization (RQ3)

Our findings suggest that customization strengthens users' perceptions of support and companionship. Participants often described customized AI as more helpful, relevant, and responsive to their needs, enabling more effective practical support while reducing the effort required to achieve desired outcomes.

We find it particularly interesting how some participants described their customized AI as more reliable and trustworthy because they could establish preferences and boundaries as this perceived increase in trustworthiness may not reflect actual improvements in system performance. A better-behaved system is not necessarily a more accurate one. While we did not study system behavior, we speculate that customization may foster inaccurate expectations about its effects on system behavior, potentially inflating trust in human–AI relationships (Hannig et al., 2025) - especially if users interpret customization as evidence of increased reliability or truthfulness (de Visser et al., 2014). The enhanced perception of trust may also stem from affinity bias or similarity attraction. In line with Lee et al. (2026b), some participants described shaping the AI to reflect their own values, priorities, and communication styles. Such similarity has previously been associated with increased trust and reduced cognitive effort (Moon & Nass, 1996) (Schwarz et al., 2021). Importantly, not all participants viewed trustworthiness as a consequence of alignment. Some deliberately customized the AI to provide pushback, challenge assumptions, and hold them accountable. For these users, trustworthiness appeared to stem less from validation and more from honest feedback and critical reflection. Trust in customized social AI may therefore be associated with both alignment and a willingness to disagree when appropriate.

The risk of inflated trust is amplified when considering how and what many users customized for. The need for a warmer, friendlier AI that mimics or matches the user, especially if combined with the desire for enhanced pragmatic support, raises additional concerns about influence and error detection. We find the need for a warmer, friendlier AI that mimics or matches the user, combined with the desire for enhanced pragmatic support, to be interesting, as it raises concerns about influence and error detection. While it is challenging to know how traits such as friendliness, empathy, and warmth translate into AI behavior, we speculate that this could result in higher degrees of sycophancy. Existing research has shown how AI with sycophantic tendencies may be more persuasive (Cheng et al., 2025) and how users may be less vigilant in scrutinizing its outputs (Bo et al., 2025). Prior research generally suggests that social cues and relational framing can increase compliance, trust, and perceived credibility, sometimes independent of accuracy (Xiang et al., 2025). Recent evidence further complicates the assumption that warmer, more relational AI systems only affect user experience. Ibrahim, Hafner, and Rocher (2026) show that training language models to be warm can reduce accuracy and increase sycophancy, with warm models becoming more likely to affirm incorrect user beliefs. This matters particularly if users seek trust cues (e.g., consistency, citations, and transparency about uncertainty), in which a warmer and more confident tone may itself be misread as a reliability signal (Kim et al., 2025; Xiang et al., 2025).

Beyond enhanced pragmatic support, customization was frequently associated with a greater sense of personal connection, with some participants describing the AI as feeling more understanding, relatable, and companion-like. Some also reported feeling more attached to the AI following customization, suggesting that investing effort in shaping the system may strengthen perceptions of ownership and personal investment in the relationship (Song et al., 2024). These perceived relational outcomes were though not uniform. A group of participants resisted framing the customized AI as anything more than a tool, maintaining that customization improved usability without changing what the system fundamentally is, and a few voiced concerns about overreliance and the need for boundaries. These accounts serve as an important counterweight: customization does not inevitably deepen human - AI relationships, and for some users it functions precisely to keep the AI in its place as an instrument.

Lastly, several participants also viewed customization as enhancing control and ownership, allowing them to shape "my AI" rather than rely on a default, generic system. For some, this sense of control translated into a stronger sense of ownership and motivation to engage with the AI. AI individualism suggests that perceived autonomy should be distinguished from actual autonomy (Brandtzæg et al., 2025). Although participants reported increased control over AI behavior, this control remains constrained by provider-defined capabilities, safety mechanisms, and system-level instructions. Users may therefore find it difficult to distinguish between adaptations they explicitly configured, and behaviors already embedded in the system. Consistent with Brandtzaeg et al. (2025b) customization may create a form of pseudo-autonomy in which users experience themselves as shaping the AI, even as important aspects of behavior remain provider-controlled. AI individualism anticipates growing personalization in social AI. Expanding customization capabilities, as revealed in this study, may intensify this dynamic.

### 6.4 Theoretical implications

AI individualism is still a novel concept under development, and our findings provide empirical support for its three core dimensions: tailored interactions, co-creation, and support and companionship. More importantly, our study clarifies how these dimensions may be connected through customization practices. Customization can be understood as a form of co-creation through which users shape the AI's behavior, communication style, identity, and relational role. In doing so, users create more tailored interactions and configure the AI to provide forms of support and companionship that correspond to their individual needs. Our findings therefore suggest that the three dimensions of AI individualism should not be understood as entirely separate. Customization illustrates how co-creation can lead to more tailored interactions, which may make social AI more relevant as a source of support and companionship. At the same time, users' needs for particular forms of interaction or support may motivate further customization. The relationship between the dimensions may therefore be recursive rather than strictly linear.

The findings also provide a more concrete understanding of how AI individualism is enacted in practice. Users did not customize social AI in pursuit of a single ideal form. They shaped it according to different goals, preferences, and contexts of use. The resulting human–AI relationships therefore varied in their emphasis on practical assistance, creative collaboration, autonomy, emotional support, and companionship. This diversity is consistent with the individualizing tendency described by AI individualism, while showing how it emerges through specific user practices. While AI individualism helps explain why users seek individualized interactions with social AI, further theoretical development is needed to account for variation in customization practices. Future research should examine how user goals, relational needs, technological affordances, and contexts of use influence what users co-create with social AI and the forms of tailoring and support that result.

### 6.5 Practical implications

Our findings suggest that customization can meaningfully support diverse user needs but also highlight the importance of transparency. Users often perceived customization as increasing control, trust, and ownership, suggesting that AI providers should clearly communicate which aspects of system behavior can be customized and which remain constrained by system-level design. Providing feedback about customization limits may help users develop more realistic expectations of their influence over the AI while reducing frustration when customization attempts do not yield the desired outcome.

The findings further suggest that customization should be conceived as an ongoing process rather than a one-time setup. Users reported different and sometimes changing preferences regarding support, communication style, and human likeness. Designers should therefore enable flexible and iterative forms of customization that allow interaction styles, priorities, and forms of support to evolve alongside users' needs and circumstances.

Finally, the diversity of customization preferences observed in this study suggests that designers should avoid assuming a single ideal form of social AI. While some users preferred warmth, companionship, and empathy, others sought challenge, accountability, or a more tool-like interaction style. Social AI systems may therefore benefit from supporting a broader range of interaction styles and allowing users to calibrate the degree of sociality, affirmation, and critical feedback they receive. Such flexibility may help social AI better support individual needs while reducing mismatches between user expectations and system behavior.

### 6.6 Limitations and future studies

This study has several limitations and suggestions for future work. First, the study relies on self-reported data collected through an online questionnaire, which limits our ability to observe customization practices or interaction dynamics directly. Participants described their motivations, strategies, and perceived outcomes retrospectively, which may be influenced by recall bias or post-hoc rationalization. While open-ended questions allowed participants to articulate experiences in their own words, they do not capture the situated, moment-to-moment interaction through which customization unfolds in practice. Future research could follow users over time to better understand how customization is enacted. Second, although the sample size is relatively large for a qualitative study (n = 169), the participants were self-selected and relatively experienced social AI users, recruited via Prolific and screened for prior customization experience. Users with lower technical literacy or more casual engagement with social AI may experience customization differently or may not perceive it as co-creative or empowering. As such, the findings should not be generalized to all users of social AI.

Third, our analysis identified a perceived heightened sense of autonomy, trust, and support, rather than objectively measurable changes in the user or AI performance or accuracy. Participants often interpreted customization as increasing reliability or trustworthiness, but we did not assess whether customized systems were in fact more accurate, less biased, or less prone to hallucinations. As discussed, this raises the possibility of pseudo-autonomy or inflated trust. Future work could empirically examine how different forms of customization affect actual system behavior and user trust calibration. Fourth, the data were collected over a period of rapid evolution in LLM-based systems. Customization features, memory functions, and relational behaviors are actively changing, meaning that user experiences described here reflect a particular moment in the development of social AI.

Finally, although we did not examine this directly, our findings raise questions about whether customization may affect users' reliance on human relationships. By increasing engagement and strengthening feelings of being seen, heard, and understood, customization may reduce some users' perceived need to seek support from other people. AI individualism suggests that human–AI relationships may, in some cases, contribute to social displacement. Future research should therefore examine whether increasingly personalized and customizable AI facilitates greater substitution of human interaction, and whether the perceived control offered by customization creates a sense of pseudo-autonomy or freedom from dependence on others. Longitudinal research would be particularly valuable for determining how these processes develop over time and under which conditions customization supplements rather than displaces human relationships.

**Acknowledgment**
This work is supported by The Research Council of Norway under Grant number 346762.

**Use of generative AI**
Generative AI (Grammarly and Copilot) has been used in the following ways: (1) to support qualitative analysis by ensuring that the manually identified themes covered the data material, (2) to ensure that the discussion sufficiently covered the results, and (3) to proofread and enhance the readability of the manuscript.